\documentstyle[12pt,twoside,fleqn,epsf,espcrc1]{article}
\newcommand{\AmS}{{\protect\the\textfont2
  A\kern-.1667em\lower.5ex\hbox{M}\kern-.125emS}}
\newcommand{\beq}{\begin{equation}}
\newcommand{\eeq}{\end{equation}}
\newcommand{\ds}{\displaystyle}
\newcommand{\beqar}{\begin{eqnarray}}
\newcommand{\eeqar}{\end{eqnarray}}
\title{
 Equilibrium and non-equilibrium effects in relativistic heavy
 ion collisions.
}
\author{
L.~V.~Bravina$^{a,b}$, E.~E.~Zabrodin$^{b,c}$,
M.~I.~Gorenstein$^{a,d}$, S.~A.~Bass$^{e}$,
M.~Belkacem$^{a}$, M.~Bleicher$^{a}$, M.~Brandstetter$^{a}$, 
C.~Ernst$^{a}$, Amand Faessler$^{c}$, W.~Greiner$^{a}$, 
S.~Soff$^{a}$, H.~St{\"o}cker$^{a}$, H.~Weber$^{a}$
\\
\vspace*{.25 cm}
{\small\it
$^a$Institute for Theoretical Physics, University of Frankfurt,
D-60054 Frankfurt, Germany} \\ 
{\small\it
$^b$Institute for Nuclear Physics, Moscow State University,
119899 Moscow, Russia} \\
{\small\it
$^c$Institute for Theoretical Physics, University of 
T\"ubingen, D-72076 T\"ubingen, Germany} \\ 
{\small\it
$^d$Bogolyubov Institute for Theoretical Physics, Kiev, Ukraine} \\
{\small\it
$^e$Department of Physics, Duke University, Durham, 
NC 27708-0305, USA} 
}

\date{}

\begin{document}
\maketitle

\begin{abstract}
{\small 
The hypothesis of local equilibrium (LE) in relativistic heavy ion 
collisions at energies from AGS to RHIC is
checked in the microscopic transport model. 
We find that kinetic, thermal, and chemical equilibration of the 
expanding hadronic matter is nearly reached in central
collisions at AGS energy for $t \geq 10$ fm/$c$ in a central cell.
At these times the equation of state may be approximated by a 
simple dependence $P \cong (0.12-0.15)\, \varepsilon$.
Increasing deviations of the yields and the energy spectra of hadrons 
from statistical model values are observed for increasing bombarding
energies. 
The origin of these deviations is traced to the irreversible 
multiparticle decays of strings and many-body $(N \geq 3)$ decays of 
resonances.
The violations of LE indicate that the matter in the cell reaches
a steady state instead of idealized equilibrium.
The entropy density in the cell is only about 6\% smaller than
that of the equilibrium state.
}
\end{abstract}

\vspace{.6 cm}

The assumption of the creation of a locally equilibrated (LE) hadronic 
state in ultrarelativistic heavy ion collisions has been subject of 
theoretical and experimental efforts during the last decades.
Despite the long history the question remains still open.
The present analysis employs the 
Ultra-relativistic Quantum Molecular Dynamics (UrQMD) model 
\cite{UrQMD} to examine the approach to local equilibrium of 
hot and dense nuclear matter, produced in central heavy ion collisions 
at energies from AGS to SPS and RHIC. 
 
\begin{figure}[htb]
\begin{minipage}[t]{69mm}
\vspace*{-1.5cm}
\centerline{\epsfysize=75mm\epsfbox{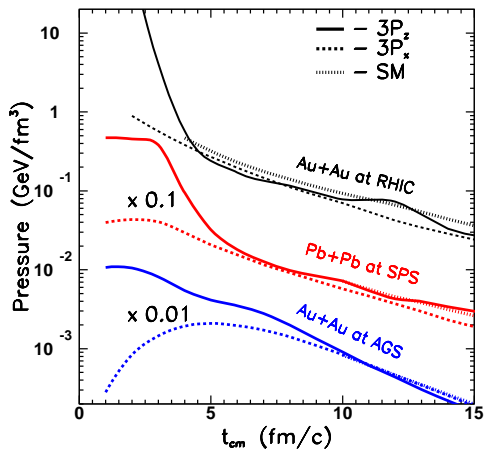}}
\vspace*{-1.4cm}
\caption{\small The longitudinal (solid) and the 
transverse (dashed) diagonal components 
of the pressure tensor $P$ in the central cell 
of heavy ion collisions compared to the SM results
(dotted lines).
 }
\label{fig1}
\end{minipage}
\hspace{\fill}
\begin{minipage}[t]{83mm}

\vspace*{-0.8cm}

Having the criterion for {\it kinetic\/} equilibrium fulfilled, one 
may address the question on {\it thermal\/} and {\it chemical\/} 
equilibrium in the cell. 
Both criteria read: {\bf (iii)} the distribution functions 
of hadrons obey Bose-Einstein or Fermi-Dirac statistics
(thermal equilibration) with the unique temperature $T$
\beq
\ds
f(p,m_i) = \left[ \exp{(\sqrt{p^2 + m_i^2} -\mu_i)/T} \pm 1 
\right]^{-1}
\label{eq2}
\eeq
(where $\mu_i$ is the chemical potential of $i$th particle,
$``+"$ sign stands for fermions and $``-"$ for bosons),
and {\bf (iv)} the yields of hadrons are calculated via $f(p,m_i)$
with $\mu_i = \mu_{\rm B}B_i + \mu_{\rm S}S_i$ (chemical equilibrium). 
The latter condition assumes that any inverse reaction 
proceeds with the {\bf same rate\/} as the direct reaction, i.e., that 
the detailed balance is fulfilled.

\end{minipage}
\end{figure}

First, the kinetic equilibration of the system is examined.
In order to diminish the number of distorting factors we choose
a cubic cell of volume $V = 125\,$fm$^3$
centered around the origin of the CM-system of the colliding nuclei.
Due to the absence of a preferential direction of the collective
motion, the collective velocity of the cell is essentially zero.
The longitudinal flow in the cell reaches its maximum value at times 
from $t = 2$ fm/$c$ (RHIC) to $t = 6$ fm/$c$ (AGS). Then it drops and
converges to the transverse flow. 
Disappearance of the flow implies:
{\bf (i)} isotropy of the velocity distributions, which leads to
{\bf (ii)} isotropy of the diagonal elements of the pressure tensor, 
calculated from the virial theorem,
\beq
\ds
P_{\{x,y,z\}} = \sum_{i=h}
p^2_{i\{x,y,z\}} /3 V (m_i^2~+~p_i^2)^{1/2} \quad,
\label{eq1}
\eeq
containing the volume of the cell $V$ and the mass and the momentum
of the $i$-th hadron, $m_i$ and $p_i$, respectively.
The time evolution of the pressure in longitudinal and transverse 
directions shows (Fig.~\ref{fig1}) that kinetic equilibration in the
central zone of the reaction takes place at $t \cong 10$ fm/$c$ (AGS),
8 fm/$c$ (SPS), and 4 fm/$c$ (RHIC).
Note that the pressure given by the SM [see Eq.~(5)] is in a good
agreement with microscopic results.
\vspace*{-0.85cm}

The standard procedure is to compare the 
snapshot of particle yields and spectra in the cell at given 
time with those predicted by the statistical thermal model of a hadron 
gas \cite{LV98plb,lv99prc}. Three parameters, 
namely the energy density $\varepsilon$, the baryon density 
$\rho_{\rm B}$, and the strangeness density $\rho_{\rm S}$, extracted 
from the analysis of the cell, are inserted into the equations for an 
equilibrated ideal gas of hadrons. Then all characteristics of the 
system in equilibrium, including the yields of different hadronic 
species, their temperature $T$, and chemical potentials, 
$\mu_{\rm B}$ and $\mu_{\rm S}$, can be calculated. 
If the yields and the energy spectra of the hadrons in the cell
are sufficiently close to those of the SM, one can take this as 
indication for the creation of equilibrated hadronic matter in the 
central reaction zone. 

The particle yields, $N_i^{\rm SM}$, and total energy, 
$E_i^{\rm SM}$, of the hadron species $i$ read
\beqar
\ds
N_i^{\rm SM} &=& \frac{V g_i}{2\pi^2\hbar^3}\int_0^{\infty}p^2
 f(p,m_i) d p \quad, \\
\label{eq3}
E_i^{\rm SM} &=& \frac{V g_i}{2\pi^2\hbar^3}\int_0^{\infty}
 p^2 \, \sqrt{p^2+m_i^2}\, f(p,m_i) d p \quad.
\label{eq4}
\eeqar
Here $g_i$ is the degeneracy factor, and the distribution function 
$f(p,m_i)$ is given by Eq.~(\ref{eq2}). 

The hadron pressure and the entropy density are calculated within 
the SM as
\beqar
\ds
P^{\rm SM}&=&\sum_i \frac{g_i}{2\pi ^2\hbar^3}\int_0^{\infty}
p^2 \frac{p^2}{3(p^2+m_i^2)^{1/2}}~f(p,m_i) d p \quad,\\
\label{eq5}
s^{\rm SM}&=&-\sum_i \frac{g_i}{2\pi^2\hbar^3} \int_0^{\infty}
f(p,m_i)\, \left[ \ln{f(p,m_i)}-1 \right] \, p^2 d p\quad.
\label{eq6}
\eeqar

\begin{figure}[htb]
\begin{minipage}[t]{70mm}
\centerline{\epsfysize=93mm\epsfbox{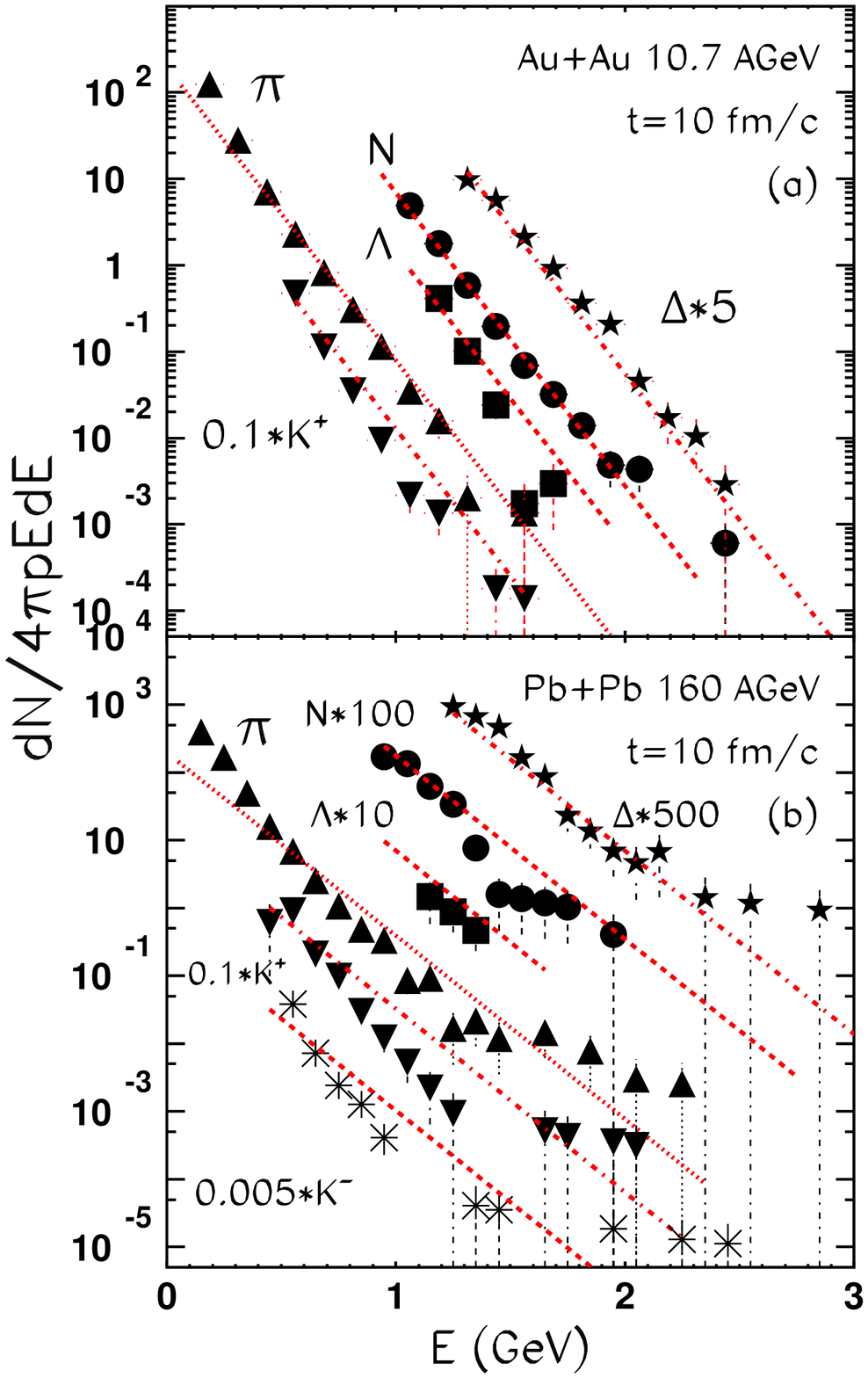}}
\vspace*{-1.1cm}
\caption{ \small
Energy spectra of hadrons in the central cell of heavy ion collisions
at AGS {\bf (a)} and SPS {\bf (b)} energies at $t$=10~fm/$c$.
Dashed lines are the results of SM fit.
    }
\label{fig2}
\end{minipage}
\hspace{\fill}
\begin{minipage}[t]{80mm}
\centerline{\epsfysize=96mm\epsfbox{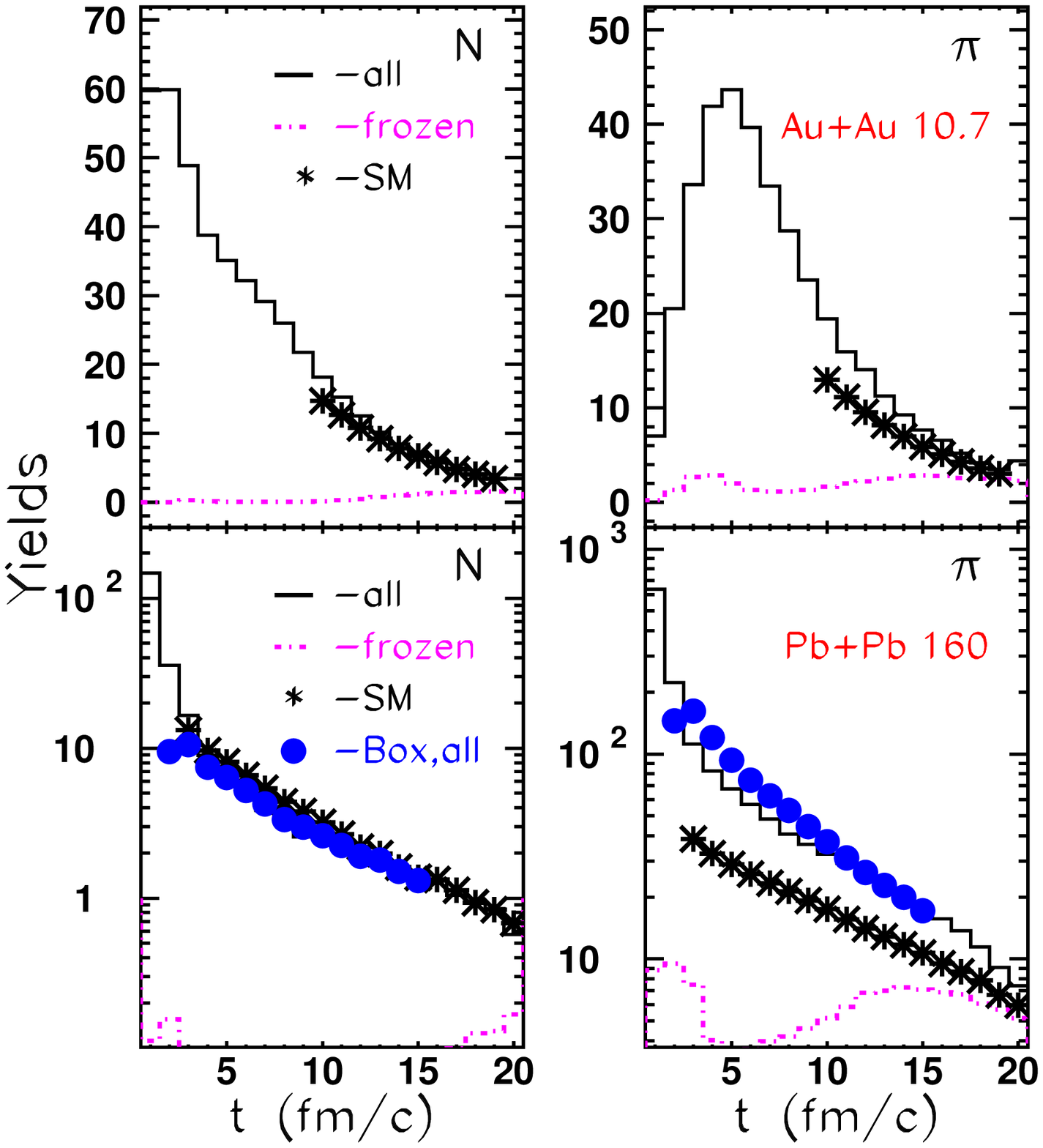}}
\vspace*{-1.1cm}
\caption{ \small
Yields of nucleons and pions in the central cell 
at AGS (upper row) and SPS (lower row) energies as a 
function of time, obtained in UrQMD model (histograms), in SM 
($\ast$), and in box calculations (circles).
  }
\label{fig3}
\end{minipage}
\end{figure}

Figure~\ref{fig2} shows the energy spectra of 
hadronic species, obtained from the microscopic calculations
together with the predictions of the SM.
At AGS energy the difference between the UrQMD
and SM results for baryons lies within the 10\%-range of accuracy.
With the rise of initial energy from AGS to SPS the agreement between
the models becomes worse. Moreover, even at 10.7 AGeV the 
deviations of pion spectra in UrQMD from those of the SM are 
significant. 

\vspace*{-1.2cm}

\vspace*{-0.9cm}
The Boltzmann fit to pion and nucleon energy spectra 
from the central cell at 160 AGeV shows \cite{lv99prc} that the 
nucleon (pion) ``temperature" is about 30 (50)~MeV below the 
$T^{\rm SM}$.
The subtraction of pions \cite{lv99prc} does not decrease 
the temperature in the SM fit, but leads to the increase of 
chemical potential of strange particles.

The yields of nucleons and pions
in the central cell are shown in Fig.~\ref{fig3}.
The agreement between the SM and UrQMD nucleon yields is reasonably 
good for $t \geq 10$ fm/$c$. Compared to UrQMD,
the statistical model significantly underestimates the number of 
pions, especially at 160 AGeV.
The conditions {\bf (iii)} and {\bf (iv)} are not satisfied.
Despite the occurrence of a state in which hadrons are
in kinetic equilibrium and collective flow is rather small, the 
hadronic matter is neither in thermal nor in chemical equilibrium.
However, the hadron multiplicities in the cell are in a good
agreement with those of the equilibrated infinite matter 
\cite{Belk98}, simulated within the UrQMD (Fig.~\ref{fig3}, circles).

\begin{figure}[htb]
\vspace*{-1.0cm}
\begin{minipage}[t]{75mm}
\centerline{\epsfysize=84mm\epsfbox{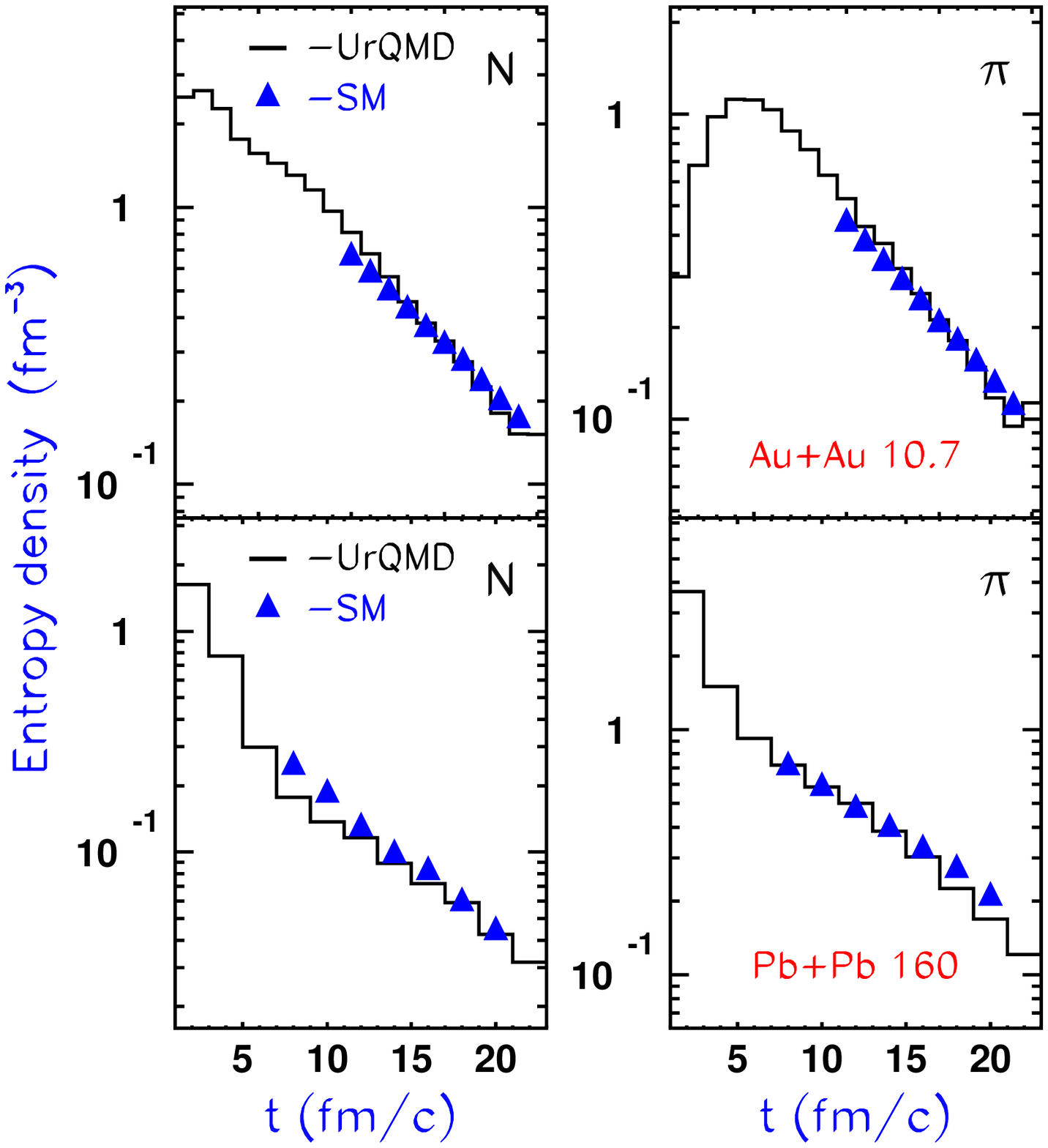}}
\vspace*{-1.1cm}
\caption{ \small
The same as Fig.~\ref{fig3} but for the entropy densities of
nucleons and pions. Histograms indicate UrQMD results, symbols
are the SM predictions.
 }
\label{fig4}
\end{minipage}
\hspace{\fill}
\begin{minipage}[t]{75mm}
\centerline{\epsfysize=84mm\epsfbox{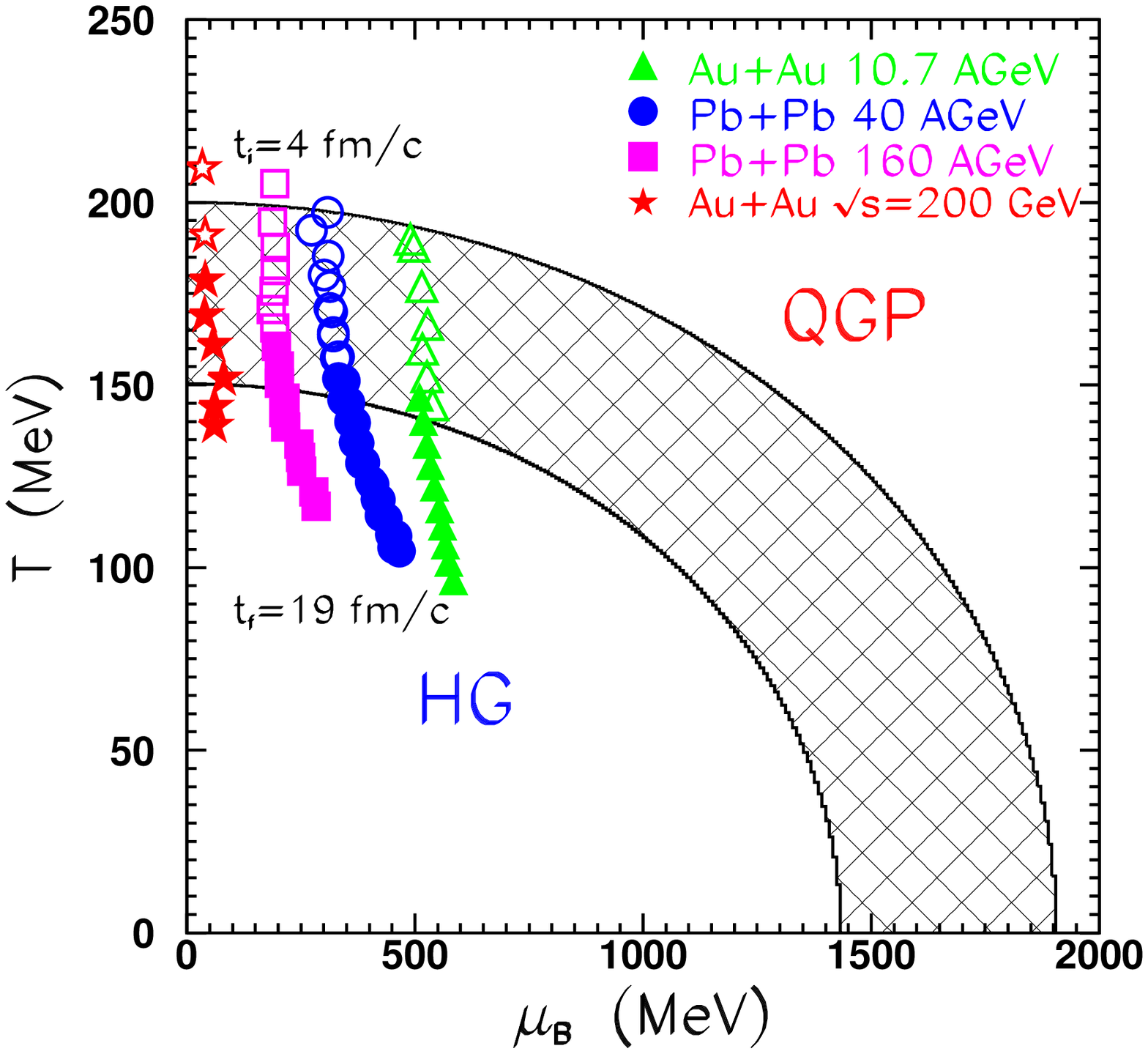}}
\vspace*{-1.1cm}
\caption{ \small
The plot $T$ vs $\mu_{\rm B}$
in the central cell of heavy ion collisions.
Open symbols correspond to the nonequilibrium stage,
full symbols indicate the stage of equilibrated
matter.
 }
\label{fig5}

\vspace*{-1.1cm}
\end{minipage}
\end{figure}

The detailed balance in relativistic heavy ion collisions is broken 
because of the irreversibility of 
multiparticle processes and non-zero lifetimes of resonances.
Thus, the matter in the cell is in a steady state
\cite{Pr61,irr99} rather than in idealized equilibrium. 
The entropy per baryon ratio stays
remarkably constant during the expansion at the quasi-equilibrium
stage \cite{lv99prc}. This fact supports the applicability of
hydrodynamics, which assumes an isentropic expansion of the
relativistic hadron liquid.
The partial entropy densities, carried separately by hadron species
in the cell (Fig.~\ref{fig4}) are close to those in the SM. The
total entropy density is only about 6\% smaller than the SM 
total entropy density.

The evolution of the central cell in $T$-$\mu_{\rm B}$ plane 
(Fig.~\ref{fig5}) indicates that the
extraction of temperature by performing the SM fit to hadron yields
and energy spectra is a very delicate procedure. Although the 
temperatures of hadrons in the steady state are limited to
$T_{lim} \leq 145$ MeV, the ``apparent" temperature
obtained from the fit may occur high enough to hit the zone of
quark-hadron phase transition or even pure QGP phase.
To study the heavy ion collisions at high energies one has to 
apply the non-equilibrium thermodynamics of irreversible processes,
and not the equilibrium thermodynamics!

\vspace*{-0.8cm}

{\bf Conclusions.} 
The results of the present study may be summarized as follows.

1.  There is a kinetic equilibrium stage of hadron-string matter in 
the central $V = 125$~fm$^3$ cell of relativistic heavy ion collisions 
at about $t \geq 8$ fm/$c$.

2. Entropy per baryon ratio remains constant during the
time interval $ 8 \leq t \leq 18$~fm/$c$. This result
supports the application of relativistic hydrodynamics.

3. The differences between the UrQMD and SM results
increase with rising bombarding energy, i.e., thermal and chemical
equilibrium is not reached. But: hadron spectra and yields in the 
cell are consistent with the UrQMD infinite matter calculations.

4. We call this quasi-equilibrium state steady state. Its origin is 
traced to the irreversible multiparticle processes and many-body 
decays of resonances.


\vspace{-0.4 cm}

\begin{thebibliography}{9}

\vspace{-0.3 cm}

{\small
\bibitem{UrQMD} 
S.A.~Bass et al., Prog. Part. Nucl. Phys. 41 (1998) 255. 

\bibitem{LV98plb} 
L.V.~Bravina et al., Phys. Lett. B 434 (1998) 379;
J. Phys. G 25 (1999) 351.

\bibitem{lv99prc} 
L.V.~Bravina et al., Phys. Rev. C, in press, preprint hep-ph/9906548.

\bibitem{Belk98} M.~Belkacem et al.,
Phys. Rev. C 58 (1998) 1727.

\bibitem{Pr61} 
I.~Prigogine,
Thermodynamics of irreversible processes, 
John Wiley \& Sons, NY, 1961.

\bibitem{irr99} 
E.E.~Zabrodin, L.V.~Bravina, H.~St\"ocker, and W.~Greiner,
preprint hep-ph/9901356.
}

\end{thebibliography}
\end{document}